\documentclass{autart}    

\usepackage{graphicx}          
\renewcommand{\thefootnote}{*\arabic{footnote}}
\usepackage{amsmath,amssymb}
\usepackage{nccmath}
\usepackage{empheq}
\usepackage{bm}
\usepackage{color}
\usepackage{blox}
\usepackage{tikz}
\usepackage[format=hang,labelsep=quad,margin=5pt]{caption}
\usepackage{multirow}
\usepackage{algorithm}
\let\classAND\AND
\let\AND\relax
\usepackage{algorithmic}

\let\AND\classAND
\AtBeginEnvironment{algorithmic}{\let\AND\algoAND}

\usetikzlibrary{calc}
\usetikzlibrary{positioning}
\usetikzlibrary{arrows.meta}

\newcommand{\del}{\partial}

\newcommand{\pdiff}[2]{\frac{\del #1}{\del #2}}

\newcommand{\PAR}[1]{{\left( #1 \right)}}

\newcommand{\PPPAR}[1]{{\left[ #1 \right]}}

\newcommand{\R}{\mathbb{R}} 
\newcommand{\T}{\top}

\newcommand{\ep}{{\varepsilon}}
\newcommand{\migi}{\rightarrow}
\newcommand{\Migi}{\Rightarrow}

\newcommand{\douti}{\Leftrightarrow}
\newcommand{\la}{\leftarrow}
\newcommand{\nn}{\nonumber}

\newcommand{\norm}[1]{{\left\| {#1} \right\|}}

\newcommand{\minimize}{\mathop{\rm minimize}\limits}

\newcommand{\mat}[1]{\begin{bmatrix}#1\end{bmatrix}}
\newcommand{\mycase}[1]{\left\{\begin{array}{ll}#1\end{array}\right.}

\newcommand{\rev}[1]{\textcolor{black}{#1}}
\newcommand{\revenv}{\color{black}}

\begin{document}

\begin{frontmatter}
\title{Sampled-Data Primal-Dual Gradient Dynamics in Model Predictive Control\thanksref{footnoteinfo}} 

\thanks[footnoteinfo]{This paper was not presented at any IFAC 
meeting. Corresponding author K.~Kashima.}

\author[tcrdl]{Ryuta Moriyasu}\ead{moriyasu@mosk.tytlabs.co.jp},    
\author[tico]{Sho Kawaguchi}\ead{sho.kawaguchi@mail.toyota-shokki.co.jp},               
\author[ku]{Kenji Kashima}\ead{kk@i.kyoto-u.ac.jp}  

\address[tcrdl]{Toyota Central R\&D Labs., Inc., Aichi, Japan}  
\address[tico]{Toyota Industries Corporation, Aichi, Japan}             
\address[ku]{Graduate School of Informatics, Kyoto University, Kyoto, 606-8501, Japan}        
          

\begin{abstract}                          
	Model Predictive Control (MPC) is a versatile approach capable of accommodating diverse control requirements that holds significant promise for a broad spectrum of industrial applications. 
	Noteworthy challenges associated with MPC include the substantial computational burden, which is sometimes considered excessive even for linear systems.
	Recently, a rapid computation method that guides the input toward convergence with the optimal control problem solution by employing primal-dual gradient (PDG) dynamics as a controller has been proposed for linear MPCs. 
	However, stability has been ensured under the assumption that the controller is a continuous-time system, leading to potential instability when the controller undergoes discretization and is implemented as a sampled-data system.
	In this paper, we propose a discrete-time dynamical controller, incorporating specific modifications to the PDG approach, and present stability conditions relevant to the resulting sampled-data system.    
	Additionally, we introduce an extension designed to enhance control performance, that was traded off in the original.
	Numerical examples substantiate that our proposed method, which can be executed in only 1~$\mu$s in a standard laptop, not only ensures stability with considering sampled-data implementation but also effectively enhances control performance.
\end{abstract}\vspace{-10pt}

\end{frontmatter}

\section{Introduction}

Model predictive control (MPC) is noteworthy as a promising control methodology in the industrial domain, offering optimal control by explicitly addressing various control requirements, including constraints.
Its applications are extensive, spanning automotive control \cite{bemporad2018model,nakada2018application,Moriyasu2019}, process control \cite{kumar2012model,lozano2016algorithm}, air conditioning \cite{afram2014theory,Drgona2018,Bunning2020}, and beyond.
Despite its potential, practical applications of MPC grapple with significant challenges, notably the formidable computational load and the intricate task of ensuring closed-loop stability \cite{rakovic2018handbook}. 
These are usually more severe when dealing with nonlinear systems, but they have not yet been completely solved for linear systems either.

MPC for linear systems has been studied extensively, and there are comprehensive formulations for stability guarantee and constraint handling.
However, the computational burden for solving linear MPC problems, which are usually reduced to convex quadratic programming problems (QPs), is still considered excessive in some applications \cite{RUAN2022123265,10534345,doi:10.1177/01423312211015120}, such as those with high-speed dynamics and limited computational resources.
In particular, embedded microcontrollers are usually much slower than processors for ordinary personal computers, and often, because of conflicts with other functions of the product, only a portion of the resource can be allocated.
\rev{The application of parallel computation is also not easy when the product design is under strong cost constraints.}
If incomplete solutions are obtained due to insufficient computation, not only control performance will be degraded, but stability and constraint fulfillment may also be compromised.
The difficulty of software verification for online optimization has also been highlighted \cite{4178103,Alessio2009}.

Various methods have been proposed to overcome these issues and are still under development.
The Explicit MPC \cite{MPC10} is a well-known method that reduces the computational load by solving a multi-parametric QP in advance to obtain a piecewise affine solution.
However, the number of piecewise domains increases significantly with the number of constraints \cite{Alessio2009}.
This results in the exhaustion of memory capacity and makes it difficult to identify the piecewise domain online.
Although QP solvers, which are primarily intended for application to MPC, have also been developed in recent years \cite{osqp,fbrs}, achieving drastic speedups appears challenging. 
There are other approaches---e.g., approximate MPC \cite{parisini1995receding}, continuation method \cite{Ohtsuka2004}, time-distributed optimization (TDO) \cite{LIAOMCPHERSON2020108973}---to reduce computation time at the cost of some loss of optimality, but they predominantly focus on nonlinear systems.
TDO provides a stability guarantee and constraint fulfillment, but the reduction of the computational load is limited for linear cases.
In contrast, the approximate MPC and continuation method are still effective in terms of computational load for linear cases, but stability and constraint fulfillment are not guaranteed in general.

This study zeroes in on instant MPC (iMPC) \cite{Yoshida2019} as a well-balanced solution to the aforementioned challenges. 
iMPC leverages primal-dual gradient dynamics (PDG) \cite{FEIJER20101974,cherukuri2016asymptotic}, where the solution to an optimization problem manifests as a fixed point within continuous-time dynamics. 
The controller, based on PDG, and the dynamics of the controlled plant evolve concurrently. 
The PDG does not fully resolve the control problem at each instance; however, it ensures satisfaction of the KKT condition of the original optimal control problem at the equilibrium point of the closed-loop system. 
This methodology facilitates rapid computations by obtaining control inputs through straightforward calculations of the time evolution of ordinary differential equations.
Moreover, closed-loop stability can be guaranteed by considering the dissipative nature of the controller and plant.

\begin{figure}[t]
	\centering
	\includegraphics[clip, scale=0.3]{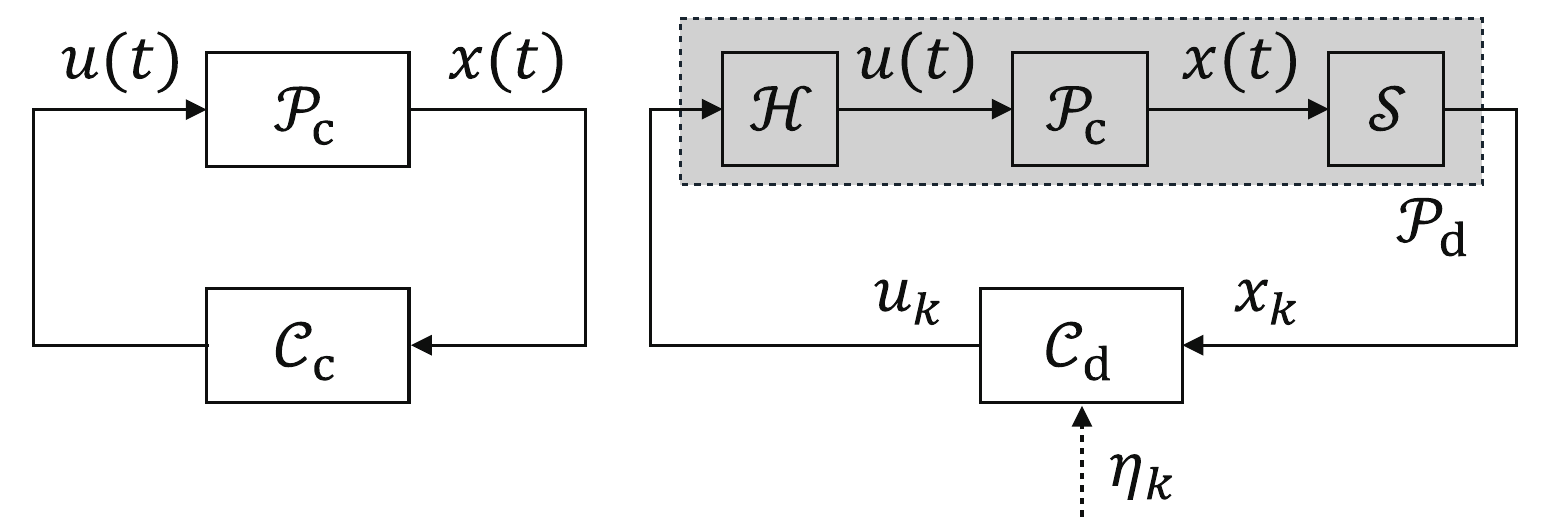}\\
	{\scriptsize (a) Continuous-time \hspace{40pt} (b) Sampled-data \hspace{20pt}  }
	\caption{Schematics of feedback systems. $\mathcal{P},\mathcal{C},\mathcal{S},\mathcal{H}$ represent a plant, a controller, a sampler, and a holder, respectively. The subscript $\rm c,d$ indicate continuous-time and discrete-time, respectively.\label{system}}
\end{figure}

Despite its merits, challenges persist regarding stability in implementation and potential performance degradation. 
The method's stability guarantee pertains to continuous-time (Figure~\ref{system}(a)), and when discretized for implementation as a sampled-data control system (Figure~\ref{system}(b)), preservation of stability is not assured, potentially leading to divergence. 
Additionally, as mentioned earlier, the method gradually converges the input to the optimal point, introducing a tendency for transient performance degradation until the closed-loop system attains equilibrium.

Considering the aforementioned challenges, we endeavor to extend the capabilities of iMPC to address these issues effectively. 
Our approach involves introducing a controller that corresponds to the discrete-time version of PDG to tackle the optimal control problem. This discrete-time controller is derived through a modified form of the naive discretization of continuous-time PDG using Eulerian forward difference. A notable distinction from the naive discretization lies in the introduction of a step size for the dual variable, denoted as $\eta$ in Figure~\ref{system}(b).

A primary contribution of this study is the derivation of stability conditions for the resulting sampled-data feedback system. 
This is achieved by explicitly considering the impact of discretization and the sampling period, leveraging a newly proposed step-size algorithm. 
Other contributions involve the proposal of introducing a gain for the controller aimed at enhancing control performance and the proposal of methods to improve constraint fulfillment while ensuring stability.

The remainder of this paper is organized as follows.
In Section~2, the problem to be solved is stated.
The structure and stability conditions of iMPC are succinctly summarized in Section 3, where we also discuss the introduction of controller gains, an aspect not covered in the original paper, to afford a more flexible control design. 
In Section 4, we present our discrete-time controller proposal and elucidate the stability conditions of the resulting sampled-data system. 
Section 5 details numerical experiments conducted under various parameter settings, evaluating stability measures based on both conventional theory and our proposed methodology. 
Furthermore, the computation efficiency of the proposed controller is demonstrated by comparing it with major solvers, including Explicit MPC and continuation method.
The concluding Section 6 provides a summary of our findings and their implications.
\vspace{-15pt}

\paragraph*{Notation}
Let us denote the set of all positive (nonnegative) vectors in $\R^n$ by $\R^n_{>0} \ (\R^n_{\geq 0})$; the set of all nonnegative integers by $\mathbb{Z}_{\geq 0}$; 
the identity matrix of appropriate size by $I$; 
the $i$-th element of a vector $v$ by $(v)_i$; 
the symmetric part of a matrix $M$ by ${\rm Sym}(M)$; 
a collective vector $[v_1^\T \cdots v_n^\T]^\T$ by $[v_1;\cdots;v_n]$; Hadamard product by $\circ$; Euclidean norm by $\norm{\cdot}$; a weighted vector norm $(x^\T P x)^{1/2}$ with positive definite symmetric matrix $P$ by $\norm{x}_P$;
 and the maximum eigenvalue of a symmetric matrix $S$ by $\lambda_{\rm max} (S)$.
We define an operator
\begin{align*}
    [a]_b^+ := \mycase{
	a & (b > 0) \\
	\max(0, a) & (b = 0)
	}.
\end{align*}
For vectors $a\in \R^n, b \in \R^n_{\geq 0}$, $[a]_b^+$ imply the vector whose $i$-th component is $[(a)_i]_{(b)_i}^+$.

\section{Problem Statement}

We consider a continuous-time linear system $\mathcal{P}_{\rm c}$:
\begin{align}
	\dot{x} &= A_{\rm c} x + B_{\rm c} u, \label{plant} 
\end{align}
where $x \in \R^{n}, u \in \R^{m}$ are the state and input, respectively, and $A_{\rm c},B_{\rm c}$ are constant matrices of appropriate dimensions.

Utilizing the discretized model $x_{k+1} = A_h x_k + B_h u_k$, where $A_h =e^{A_{\rm c}\Delta \tau}, B_h=\int_0^{\Delta \tau} e^{A_{\rm c}(\Delta \tau-t)} B_{\rm c} {\rm d}t$ and $\Delta \tau \in \R_{>0}$ is a discrete-time step, we can consider a finite-horizon optimal control problem
\begin{prob} (Optimal control problem)\label{ocp}
	\begin{align*}
	&\minimize_w \ f(w) \ \ {\rm s.t. } \ g(w) \leq  0, \\
	& h(w;x) = \mat{x_1 - (A_h x_0 + B_h u_0) \vspace{-3pt} \\\vdots \vspace{-3pt}\\ x_N - (A_h x_{N-1} + B_h u_{N-1})} = 0,
	\end{align*}
    where $f,g,h$ are the objective function, inequality constraint function, and equality constraint function, respectively, and $w:=[u_{0:N-1};x_{1:N}] \in \R^{(n+m)N}$ is a combined vector consisting of input sequence $u_{0:N-1}:=[u_0;\ldots;u_{N-1}]\in \R^{mN}$ and state sequence $x_{1:N}:=[ x_1;\ldots;x_{N}]\in \R^{nN}$ with horizon length $N \in \mathbb{N}$ and initial state $x_0 = x$.
\end{prob}
Here, $h(w;x)$ can additionally include any linear equality constraint.

\section{Preliminaries}

The PDG-based MPC controller named instant MPC (iMPC) \cite{Yoshida2019} and the theory of stability guarantee are outlined here, with certain extensions to the original.

\subsection{Assumptions}

We assume that the plant is dissipative with quadratic supply rate (QSR-dissipativity) \cite{willems1972dissipative,rahnama2016}~:
\begin{assum} \label{QSR}
	(Assumption for the plant)\\
	For $S^{\mathcal{P}}(x) := \frac{1}{2} x^\T x$, there exists $Q=Q^\T \prec 0, \, S,\, R=R^\T$ such that 
	\begin{align*}
		\dot{S}^{\mathcal{P}}\leq \mat{x\\u}^\T H^{\mathcal{P}}_{\rm c} \mat{x\\u}, \
		H^{\mathcal{P}}_{\rm c} := \mat{Q & S\\S^\T & R} 
	\end{align*}
	holds for any $x,u$.
\end{assum}
\begin{rem}
	This assumption can be satisfied with matrices such as
	\begin{align*}
		H^{\mathcal{P}}_{\rm c} &= \mat{ {\rm Sym}(A_{\rm c}) & B_{\rm d}/2 \\ B_{\rm d}^\T/2 & O }.
	\end{align*}
    if ${\rm Sym}(A_{\rm c}) \prec 0$.
	Even if not, you can design a pre-controller $u=Kx+v$, with new input $v$ and gain $K$, for any controllable linear systems to satisfy ${\rm Sym}(A_{\rm c}+B_{\rm c}K) \prec 0$ and just replace $A_{\rm c}$ and $u$ by $A_{\rm c}+B_{\rm c}K$ and $v$. 
\end{rem}

We assume Problem~\ref{ocp} to exhibit the following properties:
\begin{assum} \label{ass_iMPC1}
	(Assumption for the control problem)
	\begin{enumerate}
		\renewcommand{\labelenumi}{\alph{enumi})}
		\item $f$ is $\sigma$--strongly convex and satisfies $\nabla f(0) = 0$. 
		\item $g$ is convex and satisfies $g(0) \leq 0$.
		\item $h$ is affine as $h(w;x) = Cw + Dx$.
	\end{enumerate}
\end{assum}
The operator $\nabla$ shows the partial derivative concerning $w$ hereafter. 
Note that (a) yields $(\nabla f(w))^\T w \geq \sigma w^\T w$, which is originally assumed instead of strong convexity in \cite{Yoshida2019}.

\subsection{Instant MPC}

A continuous-time dynamics $\mathcal{C}_{\rm c}:$
\begin{align*}
    \dot{w} &= -\zeta \PAR{ \nabla f(w) \!+\! \nabla g(w) \mu \!+\! \kappa \nabla h(w;x)
		 \PAR{ \lambda \!+\! \frac{\beta}{\zeta} \dot{\lambda}}}, \\
	\dot{\mu} &= \zeta \PPPAR{ g(w) }_\mu^+ ,\\
	\dot{\lambda} &= \zeta \tau \PAR{ -\alpha \lambda + h(w;x) },\\
	u &= Ew,
\end{align*}
called iMPC provides input close to the solution of Problem~\ref{ocp}.
Here, $\mu \in \R^{n_\mu},\lambda\in \R^{n_\lambda}$ are the dual variables corresponding to the inequality and equality constraints, respectively, $\alpha \in \R_{>0},\beta \in \R_{\geq 0}, \zeta \in \R_{>0}$ are scalar constants. 
$E$ is a constant matrix satisfying $Ew=u_0$, and $\kappa := 1+2\alpha \beta,\tau:=(1+\alpha\beta)^{-1}$.
The proposed controller can be viewed as a variation of PDG dynamics \cite{cherukuri2016asymptotic}, incorporating a slight modification to enable adjustable dissipativity by introducing $\alpha,\beta$.

The parameter $\zeta$ introduced here is absent in the original study \cite{Yoshida2019}. 
This constant gain serves as an essential factor for tuning the overall response speed of the controller. 
Subsequently, in the ensuing propositions, adjustments are made to the results presented in the original paper to accommodate the inclusion of the parameter $\zeta$ into the framework.

\begin{rem}\label{iMPC_remark}
	Assuming that $x$ is fixed at some point, the constant gain $\zeta$ can adjust the speed to reach the fixed point of $\mathcal{C}_{\rm c}$:
	\begin{align}
		&\nabla f(w) + \nabla g(w) \mu + \kappa\nabla h(w;x) \lambda =0, \nn \\
        &\PPPAR{ g(w) }_\mu^+ = 0, h(w;x) = \alpha \lambda.\label{iMPC_stab}
	\end{align}
	This condition does not exactly correspond to the KKT condition of Problem~\ref{ocp}:
	\begin{align}
	    &\nabla f(w) + \nabla g(w) \mu + \nabla h(w;x) \lambda =0, \nn \\
        &\PPPAR{ g(w) }_\mu^+ = 0, h(w;x) = 0.\label{KKT}
	\end{align}
	This observation implies that even when the input satisfies the equilibrium condition (\ref{iMPC_stab}) corresponding to $x$ at each time, the resultant trajectory of the feedback system $(\mathcal{P}_{\rm c},\mathcal{C}_{\rm c})$ may not align with the solution to Problem~\ref{ocp}.
	Nevertheless, in the event that the origin $(w,\mu,\lambda,x)=(0,0,0,0)$ of the feedback system is asymptotically stable, the point corresponding to (\ref{iMPC_stab}) asymptotically converges to the KKT point (\ref{KKT}).
\end{rem}

The subsequent results pertain to the stability analysis of the continuous-time feedback system $(\mathcal{P}_{\rm c},\mathcal{C}_{\rm c})$.
\begin{lem}\cite{Yoshida2019} \label{cont_dissipativity}
	Let $S^{\mathcal{C}}(w,\mu,\lambda):=\frac{1}{2}(w^\T w + \mu^\T \mu + \lambda^T \lambda)$ be the storage function of $\mathcal{C}_{\rm c}$. 
	Under Assumption~\ref{ass_iMPC1}, the dissipation inequality 
	\begin{align*}
	\dot{S}^{\mathcal{C}}\leq
	 \zeta \begin{bmatrix}
		w\\
		x
	 \end{bmatrix}^\T
	 H^{\mathcal{C}}_{\rm c}
	 \begin{bmatrix}
		w\\
		x
	 \end{bmatrix},
	 H^{\mathcal{C}}_{\rm c} :=
	 \begin{bmatrix}
		-\sigma I -\beta C^\T C & -\beta C^\T D\\
		-\beta D^\T C & \frac{\tau}{4\alpha}D^\T D
	 \end{bmatrix}
	\end{align*}
	holds for any $w,x$\footnote{If $f(w)=\frac{1}{2}\norm{w}^2_P \ (P=P^\T,\sigma I \preceq P)$, then $\sigma I$ can be replaced as $P$ hereafter to reduce conservatism.}.
\end{lem}

\begin{prop}\label{iMPC_stability}\cite{Yoshida2019}
	Under Assumptions~\ref{QSR} and \ref{ass_iMPC1}, if there exists $\delta \in \R_{>0}$ satisfying
	\begin{align*}
		H_{\rm c} &:= H^{\mathcal{C}}_{\rm c} + \delta W^\T H^{\mathcal{P}}_{\rm c} W \
        \prec 0, \ W := \mat{O & I \\ E & O}
	\end{align*}
	then $(w,\lambda,x)=(0,0,0)$ is globally asymptotically stable and $(w,\mu,\lambda,x)=(0,0,0,0)$ is Lyapunov stable.
\end{prop}
Note that $W$ is a matrix that converts $[x;u]$ into $[w;x]$ by $u=Ew$, i.e., $[x;u] = W [w;x]$.
This proposition implies that
\begin{align}
	V(w,\mu,\lambda,x) := S^{\mathcal{C}}(w,\mu,\lambda) + \delta \zeta S^{\mathcal{P}}(x) \ (\geq 0) \label{Lyap}
\end{align}
is a Lyapunov function.

\subsection{Additional Considerations}

The presented Proposition~\ref{iMPC_stability} stands as the central outcome in the literature~\cite{Yoshida2019}. 
Asymptotic stability of the dual variable $\mu$ is not claimed in the result.
However, it can be feasibly demonstrated by introducing a slight refinement to Assumption~\ref{ass_iMPC1}, as follows:
\begin{assum} \label{ass_iMPC2}
	(Slightly tightened assumption for the control problem)\
	\begin{enumerate}
		\renewcommand{\labelenumi}{\alph{enumi})}
		\item $f$ is $\sigma$--strongly convex and satisfies $\nabla f(0) = 0$. 
		\item $g$ is convex and satisfies $g(0) < 0$.
		\item $h$ is affine as $h(w;x) = Cw + Dx$.
	\end{enumerate}
\end{assum}
\noindent
The modification to Assumption \ref{ass_iMPC1}(b) involves eliminating the equality for the inequality function $g$. 
This assumption \ref{ass_iMPC1}(b) necessitates that $w=0$ serves as an interior point within the feasible set associated with the inequality constraint $g(w)\leq0$. 
The implication of this adjustment is encapsulated in the subsequent result.

\begin{thm}\label{iMPC_full_stability}
	Under Assumption~\ref{QSR} and \ref{ass_iMPC2}, if there exists $\delta$ satisfying $H_{\rm c}\prec 0$, then $(w,\mu,\lambda,x)=(0,0,0,0)$ is globally asymptotically stable.
\end{thm}
\begin{pf}
	Let (\ref{Lyap}) be a Lyapunov function candidate, then $\dot{V} = \dot{S}^{\mathcal{C}}+\delta \zeta \dot{S}^{\mathcal{P}}\leq -\zeta \norm{[w;x]}^2_{-H_{\rm c}} \leq 0$ 
	holds from $H_{\rm c}\prec 0$.

	The following shows that the equality is satisfied ($\dot{V}=0$) if and only if $(w,\mu,\lambda,x)=(0,0,0,0)$.
	When $(w,x)\neq(0,0)$, the equality fails trivially.
	When $(w,x)=(0,0)$, equality fails from 
	\begin{align*}
		\dot{V} &= \mu^\T \dot{\mu} + \lambda^\T \dot{\lambda} = 
		\zeta (\mu^\T \PPPAR{g(0)}_\mu^+ - \tau\alpha \lambda^\T \lambda) \\
		&\leq -\zeta \tau \alpha \lambda^\T \lambda \leq 0
	\end{align*}
	when $\lambda \neq 0$.
	When $(w,\lambda,x)=(0,0,0)$ and $\mu\neq 0$, the equality fails from $\dot{V} = \zeta \mu^\T \PPPAR{g(0)}_\mu^+ = \zeta \mu^\T g(0) < 0$.

	Therefore, the equality holds only if $(w,\mu,\lambda,x)=(0,0,0,0)$.
	The sufficient condition $(w,\mu,\lambda,x)=(0,0,0,0) \allowbreak \Migi \dot{V}=0$ holds trivially.
\end{pf}

Note that the result of Proposition~\ref{iMPC_stability} and Theorem~\ref{iMPC_full_stability} remains independent of the controller gain $\zeta$. 
This signifies that the stability of the continuous-time feedback system $(\mathcal{P}_{\rm c},\mathcal{C}_{\rm c})$ remains invariant, irrespective of arbitrary adjustments to the response speed of the controller. 
While the inclusion of the gain enhances control performance, an increase in the gain could have a substantial impact on stability when the controller is implemented as a sampled-data system. 
Therefore, we need to quantitatively assess the effect of discretization on stability.

The following lemma clarifies the results of the next section.
\begin{lem}\label{iMPC_fullmatrix}
	$H_{\rm c}\prec 0$ is equivalent to 
	\begin{align*}
		\bar{H}_{\rm c} :=
		 \bar{H}^{\mathcal{C}}_{\rm c} + \delta \bar{W}^\T H^{\mathcal{P}}_{\rm c} \bar{W} \prec 0,
	\end{align*}
	where $\bar{W} := [W | O]$ (i.e. $[x;u] = \bar{W} [w;x;\lambda]$) and
	\begin{align}
		\bar{H}^{\mathcal{C}}_{\rm c} &:= \mat{-\sigma I \! -\! \tau \kappa \beta C^\T C & -\frac{\tau\kappa\beta}{2} C^\T D & -\tau \alpha \beta C^\T \\
		-\frac{\tau\kappa\beta}{2}D^\T C & O & \frac{\tau}{2} D^\T \\
	    -\tau \alpha \beta C & \frac{\tau}{2} D & -\tau\alpha I}. \label{HbarCc}
	\end{align}
\end{lem}
\begin{pf}
	Since $-\tau\alpha I\prec 0$, $\bar{H}_{\rm c} \prec 0$ is equivalent to the condition for the Schur complement
	\begin{align*}
		&\mat{-\sigma I -\tau \kappa \beta C^\T C & -\frac{\tau\kappa\beta}{2} C^\T D \\
		-\frac{\tau\kappa\beta}{2}D^\T C & O } + \delta W^\T H^{\mathcal{P}}_{\rm c} W \nn\\
        &- 
		\mat{ -\tau \alpha \beta C^\T \\
		 \frac{\tau}{2} D^\T } (-\tau \alpha I)^{-1} 
		 \mat{-\tau \alpha \beta C & \frac{\tau}{2} D } \nn \\
		&= H^{\mathcal{C}}_{\rm c} + \delta W^\T H^{\mathcal{P}}_{\rm c} W = H_{\rm c} \prec 0.
	\end{align*}
\end{pf}

\section{Method}

In this section, we introduce a model predictive controller based on discrete-time PDG, which ensures stability while treating the feedback system as a sampled-data system.

\subsection{Sampled-data System}

We examine the sampled-data system derived from the system stated in Section~3, illustrated in Figure~\ref{system}(b), which comprises a sampler $\mathcal{S}$ with a sampling period $\Delta t\in \R_{>0}$ and a zero-order holder $\mathcal{H}$. 
This configuration is described by $x_k = x(k\Delta t)$ and $u(k\Delta t+d) = u_k \ (d\in[0,\Delta t), k\in \mathbb{Z}_{\geq 0})$. 
Subsequently, the cascaded connection of $\mathcal{H},\mathcal{P}_{\rm c},\mathcal{S}$ can be precisely transformed into a discrete-time linear system $\mathcal{P}_{\rm d}$:
\begin{align*}
	x_{k+1} &= A_{\rm d} x_k + B_{\rm d} u_k, 
\end{align*}
where $A_{\rm d}=e^{A_{\rm c}\Delta t}, B_{\rm d}=\int_0^{\Delta t} e^{A_{\rm c}(\Delta t-\tau)} B_{\rm c} d\tau$.
Consequently, ensuring the stability of the sampled-data feedback system is equivalent to ensuring the stability of the discrete-time feedback system $(\mathcal{P}_{\rm d},\mathcal{C}_{\rm d})$.

\subsection{Assumptions}

We assume that the plant satisfies

\begin{assum} \label{ass_diMPC_plant}
	There exists matrix $H^{\mathcal{P}}_{\rm d}$, with negative-definite upper-left $n\times n$ block,  and positive semi-definite matrix $P^{\mathcal{P}}_{\rm d}$ such that 
	\begin{align}
		\Delta S^{\mathcal{P}}_k &:= S^{\mathcal{P}}(x_{k+1}) - S^{\mathcal{P}}(x_{k}) \nn \\
		& \leq \Delta t \mat{x_k \\u_k}^\T \PAR{ H^{\mathcal{P}}_{\rm d} + \Delta t P^{\mathcal{P}}_{\rm d} }\mat{x_k\\u_k} \label{ineq_diMPC_plant}
	\end{align}
	holds for any $x_k,u_k$.
\end{assum}
\begin{rem}\label{rem_dimpc_plant}
	Assumption~\ref{ass_diMPC_plant} is an extension of Assumption~\ref{ass_iMPC1}.
    If ${\rm Sym}(A_{\rm c}) \prec 0$, this is satisfied by, e.g.,
	\begin{align*}
		H^{\mathcal{P}}_{\rm d} &= \frac{1}{\Delta t}\mat{{\rm Sym}(A_{\rm d}) -I
		 & B_{\rm d}/2 \\ B_{\rm d}^\T/2 & O }, \\
		P^{\mathcal{P}}_{\rm d} &= \frac{1}{2 (\Delta t)^2} \mat{A_{\rm d}-I \\ B_{\rm d}^\T} \mat{A_{\rm d}-I & B_{\rm d}}.
	\end{align*}
	At $\Delta t \migi 0$, $H^{\mathcal{P}}_{\rm d}$ converges to $H^{\mathcal{P}}_{\rm c}$ satisfying Assumption~\ref{QSR} because $(A_{\rm d}-I)/\Delta t \migi A_{\rm c}, B_{\rm d}/\Delta t \migi B_{\rm c}$.
\end{rem}

Problem~\ref{ocp} for this system is structured in a manner analogous to the previous section. 
The sampling period $\Delta t$ of the described system and the time step $\Delta \tau$ in Problem~\ref{ocp} can be distinct.
Here, alongside Assumption~\ref{ass_iMPC2}, we introduce an additional consideration: 
\begin{assum} \label{ass_diMPC_cont}
		In addition to Assumption~\ref{ass_iMPC2}, $f$ is $\rho$--smooth convex\footnote{A function $f(x)$ is called $\rho$--smooth convex function if it satisfies $\norm{\nabla{f(x)}-\nabla{f(y)}}\leq \rho \norm{x-y}$ for all $x,y$.} and $\mu_0 \geq 0$.
\end{assum}
\begin{lem} \label{lem_smooth}
	Under Assumption~\ref{ass_diMPC_cont}, there exists a positive semi-definite matrix $\bar{P}$ satisfying
	\begin{align}
		&\norm{\nabla f + \tau \kappa \nabla h \PAR{\lambda_k + \beta h }}^2
		\leq z_k^\T \bar{P} z_k, \label{ineq_lem_smooth}
	\end{align}
	where $z_k:=[w_k;x_k;\lambda_k]$.
\end{lem}
\begin{pf}
	We can denote $\tau \kappa \nabla h \PAR{\lambda_k + \beta h } = X z_k$ with
	$X:= \tau \kappa C^\T\PPPAR{\beta C \ \beta D \ I}$.
	Here, $\norm{\nabla f(w_k)}\leq \rho \norm{w_k}$ holds from $\rho$--smooth convexity and $\nabla f(0)=0$, and this yields
	\begin{align*}
		\norm{\nabla f \!+\! X z_k }^2 &\leq \rho^2 \norm{w_k}^2 \!+\! 2 \rho \norm{w_k} \norm{X z_k} \!+\! z_k^\T X^\T X z_k \\
		& \leq z_k^\T \bar{P} z_k
	\end{align*}
	with $\bar{P}=(\rho^2+2\rho \norm{X})I + {X^\T X}$.
\end{pf}
\begin{rem}
	A quadratic objective $f(w)=\frac{1}{2}\norm{w}_P^2 \ (P=P^\T, \sigma I \preceq P\preceq \rho I)$ is $\rho$--smooth and $\sigma$--strongly convex. For this function, the inequality in the above proof can be replaced as $\norm{\nabla f + X z_k }^2 = \norm{Pw_k + Xz_k}^2 = z_k^\T \bar{P} z_k, \ \bar{P}= \PAR{\PPPAR{P \ O \ O} + X}^\T \PAR{\PPPAR{P \ O \ O} + X}$.
\end{rem}

\subsection{Controller Design}

We consider a discrete-time dynamical controller $\mathcal{C}_{\rm d}$:
\begin{align*}
    w_{k+1} &= w_k + \Delta w_k, \\
	\mu_{k+1} &= \mu_k + \Delta \mu_k,\\
	\lambda_{k+1} &= \lambda_k + \Delta \lambda_k ,\\
 	\Delta w_k &:= -\zeta \Delta t \PAR{ \nabla f \!+\! \nabla g (\eta_k \!\circ\! \mu_k) \!+\! \kappa \nabla h\PAR{ \lambda_k \!+\! \frac{\beta \Delta \lambda_k}{\zeta\Delta t} } },\\
	\Delta \mu_k &:= \zeta \Delta t \eta_k \circ \PPPAR{g(w_k)}_{\mu_k}^+,\\
	\Delta \lambda_k &:= \zeta \tau \Delta t \PAR{ -\alpha \lambda_k + h(w_k;x_k) },\\
	u_k &= E w_k ,
\end{align*}
where $\eta$ is a step-size vector of the same dimension as $\mu$.
If $\eta$ is set to a fixed unit vector, this controller aligns with a discretization of $\mathcal{C}_{\rm c}$ using Eulerian forward differences. 
In other words, the primary distinction between the naive implementation of iMPC and our proposed system lies in how the step size $\eta_k$ (depicted in Figure~\ref{system}(b)) is determined. 
Here we show the definition of $\eta_k$ as:

\begin{defn}{(Step size vector)}\label{def_stepsize}
	\begin{align*}
        \eta_k &:= \gamma_k \bar{\eta}_k,\\
		 (\bar{\eta}_k)_i &:=
			\! \mycase{1 \ \PAR{ {\rm if } \ (\mu_k)_i \! +\! \zeta \Delta t (\PPPAR{g(w_k)}_{\mu_k}^+)_i \! \geq \! 0 } \\
					-(\mu_k)_i / \zeta \Delta t (\PPPAR{g(w_k)}_{\mu_k}^+)_i \ \ \PAR{\rm otherwise} }\! \!,\nn 
	\end{align*}
	where $\gamma_k\in (0,1]$ is a time-varying scalar. 
\end{defn}
The time-varying scalar $\gamma_k$ is determined each time to ensure stability in the manner presented later.
We can confirm that $(\eta_k)_i \geq\! \PAR{>} \ 0, \ (\mu_{k+1})_i \geq 0$ holds if $(\mu_k)_i \geq\! (>) \ 0 $.
Therefore, under Assumption~\ref{ass_diMPC_cont}, the non-negativity of $\mu_k$ is ensured at any time.

We examine the stability of the closed-loop system $(\mathcal{P}_{\rm d},\mathcal{C}_{\rm d})$ below.
First, we show the discrete-time version of Lemma~\ref{cont_dissipativity}. Proof is stated in the Appendix~1.
\begin{lem} \label{diMPC_cont}
	Under Assumption~\ref{ass_diMPC_cont}, there exists functions $a(\mu,z), \, b(\mu,z)$ satisfying 
	\begin{align}
		&\Delta S^{\mathcal{C}}_k := S^{\mathcal{C}}(w_{k+1},\mu_{k+1},\lambda_{k+1}) - S^{\mathcal{C}}(w_{k},\mu_{k},\lambda_{k}) \label{ineq_diMPC_cont}\\
		& \leq \zeta \Delta t \PAR{ z_k^\T \PAR{ \bar{H}^{\mathcal{C}}_{\rm c} \!+\! \zeta\Delta t \bar{P}^{\mathcal{C}}_{\rm d} } z_k \!+\! a(\mu_k,z_k)\gamma_k^2 \!+\! b(\mu_k,z_k) \gamma_k },	\nn
	\end{align}
	$a(0,0)=0, b(0,z) = 0 \ \forall z, b(\mu,0)<0 \ \forall \mu \neq0$, where 
	\begin{align}
		\bar{P}^{\mathcal{C}}_{\rm d} := \frac{1}{2} \PAR{\bar{P} + \tau^2 \PPPAR{C \ D \ -\alpha I}^\T \PPPAR{C \ D \ -\alpha I} }. \label{PbarCd}
	\end{align}
\end{lem}

Here, we define the discrete-time step change of Lyapunov function candidate $V$ as 
\begin{align*}
	\Delta V_k:= V(w_{k+1},\mu_{k+1},\lambda_{k+1},x_{k+1}) - V(w_{k},\mu_{k},\lambda_{k},x_{k}).
\end{align*}
We can show the closed-loop stability by employing $\gamma_k$ that achieves a monotonic decrease of $\Delta V_k$, which always exists under the condition regarding matrices $\bar{H}^{\mathcal{C}}_{\rm c}, \bar{P}^{\mathcal{C}}_{\rm d}$ (see (\ref{HbarCc}),(\ref{PbarCd})) and $H^{\mathcal{P}}_{\rm d}, P^{\mathcal{P}}_{\rm d}$ (see (\ref{ineq_diMPC_plant})).
\begin{thm} \label{thm_stability}
	Under Assumptions~\ref{ass_diMPC_plant} and \ref{ass_diMPC_cont}, 
	if
	\begin{align}
		\bar{H}_{\rm d} := \bar{H}^{\mathcal{C}}_{\rm c} +  \zeta \Delta t \bar{P}^{\mathcal{C}}_{\rm d} + \delta \bar{W}^\T ( H^{\mathcal{P}}_{\rm d} + \Delta t P^{\mathcal{P}}_{\rm d} ) \bar{W}  \prec 0  \label{cond_d}
	\end{align}
	holds with $\delta > 0$, then 
	\begin{enumerate}
		\renewcommand{\labelenumi}{\alph{enumi})}
		\item There always exists a positive scalar $\bar{\gamma}_k > 0$ such that $\Delta V_k <(=)\ 0$ holds for all $\gamma_k \in (0,\bar{\gamma}_k)$. 
		The equality $(=)$ holds iff $(w_k,\mu_k,\lambda_k,x_k)= (0,0,0,0)$.
		\item The origin $(w,\mu,\lambda,x)=(0,0,0,0)$ of the discrete-time feedback system $(\mathcal{P}_{\rm d},\mathcal{C}_{\rm d})$ is globally asymptotically stable by choosing $\gamma_k$ satisfying (a) for all $k$.
	\end{enumerate}
\end{thm}
\begin{pf} 
	Since (b) is trivial under (a), we only prove (a).
	Combining two inequalities (\ref{ineq_diMPC_plant}) and (\ref{ineq_diMPC_cont}) yields 
	\begin{align}
		\Delta V_k &= \Delta S^{\mathcal{C}}_k + \zeta \delta \Delta S^{\mathcal{P}}_k \label{pf_dVineq}\\
		&\leq \zeta \Delta t \PAR{ a(\mu_k,z_k) \gamma_k^2 + b(\mu_k,z_k)\gamma_k -\norm{z_k}^2_{-\bar{H}_{\rm d}} }.\nn
	\end{align}
	If $z_k \neq 0$, small $\bar{\gamma}_k$ satisfies (a) because $-\norm{z_k}^2_{-\bar{H}_{\rm d}}<0$.
	If $z_k = 0$ and $\mu_k \neq 0$, $\Delta V_k \leq \zeta \Delta t \PAR{ a(\mu_k,0) \gamma_k + b(\mu_k,0) } \gamma_k$ holds from (\ref{pf_dVineq}) and small $\bar{\gamma}_k$ satisfies (a) because $b(\mu_k,0) < 0$.
	If $z_k = 0$ and $\mu_k = 0$, this means $(w_k,\mu_k,\lambda_k,x_k)= (0,0,0,0)$ and (a) is trivial because $\Delta V_k = 0\ \forall \ \gamma_k \in (0,1]$.
\end{pf}
To summarize what is required to guarantee stability, we need to find $\delta>0$ that satisfies (\ref{cond_d}) and determine $\gamma_k$ satisfying (a) at each time step.
The existence of $\delta$ can be confirmed by solving an LMI, e.g., $\min_{\{\delta,\ep\}} \ep \ \text{s.t. } \ep I \succeq \bar{H}_{\rm d} $.
We can see from (\ref{pf_dVineq}) that $\gamma_k$ can be determined by $\gamma_k = \min( 1, c \bar{\gamma}_k),\ c\in(0,1)$ with deriving $\bar{\gamma}_k$ as the positive solution of an equation\footnote{An example of functions $a,b$ is shown in the proof of Lemma~\ref{diMPC_cont} (see Appendix). Replacing $- \norm{z_k}^2_{-\bar{H}_{\rm d}}$ as $\ep \norm{z_k}^2$ using the LMI solution $\ep$ is possible to avoid a large matrix computation.} $a(\mu_k,z_k) \gamma_k^2 + b(\mu_k,z_k)\gamma_k - \norm{z_k}^2_{-\bar{H}_{\rm d}} = 0$.
However, evaluating each term in the equation is complicated.

To achieve easier implementation, Algorithm~1 can be used to determine $\gamma_k$.
The algorithm only needs to evaluate $\Delta V$ directly by time evolving the discrete-time dynamics $\mathcal{P}_{\rm d}$ and $\mathcal{C}_{\rm d}$.
Although the algorithm involves a while loop, it is never stuck in an infinite loop because of (a).
In practice, the algorithm seldom generates even one additional iteration, and $\gamma_k=1$ is applied.
This is because the stability condition involves conservatism arising from high-dimensional matrix inequalities.

\begin{figure}[!t]
	\begin{algorithm}[H]
		\caption{Find stable step size coefficient $\gamma_k$}
		\label{alg1}
		\begin{algorithmic}[1]
		\REQUIRE $c\in(0,1), w_k, x_k, \lambda_k, \mu_k$
		\ENSURE  $\gamma_k$
		\STATE $\gamma_k \leftarrow 1$
		\IF{$(w_k,x_k,\lambda_k,\mu_k) \neq (0,0,0,0)$}
		\WHILE{$\Delta V_k \geq 0$}
		\STATE $\gamma_k \leftarrow c \gamma_k $
		\ENDWHILE
		\ENDIF
		\RETURN $\gamma_k$
		\end{algorithmic}
	\end{algorithm}
\end{figure}


Since $\bar{H}_{\rm d} \migi \bar{H}^{\mathcal{C}}_{\rm c} + \delta \bar{W}^\T H^{\mathcal{P}}_{\rm c} \bar{W} = \bar{H}_{\rm c}$ at $\Delta t \migi 0$ and $\bar{H}_{\rm c}\prec 0 \douti H_{\rm c}\prec 0$ holds from Lemma~\ref{iMPC_fullmatrix}, the condition (\ref{cond_d}) can be regarded as a discrete-time version of the stability condition $H_{\rm c}\prec 0$ in Proposition~\ref{iMPC_stability}.
Positive semi-definite matrices $\bar{P}^{\mathcal{P}}_{\rm d}, \bar{P}^{\mathcal{C}}_{\rm d}$ can be considered to indicate the degradation of stability due to the sampled-data implementation.

\begin{rem}\label{zetadeltat}
	An increase in $\zeta$ to enhance control performance necessitates a decrease in $\Delta t$ to meet the stability condition. 
	If $\Delta t$ is sufficiently small and $\zeta \gg \delta$, the effect of $P^{\mathcal{P}}_{\rm d}$ diminishes, i.e., $\bar{H}_{\rm d}\approx \bar{H}_{\rm c} + \zeta \Delta t \bar{P}^{\mathcal{C}}_{\rm d}$. 
	In such cases, if $\zeta\Delta t$ remains constant, the stability is largely unaffected.
\end{rem}

\subsection{Improving constraint fulfillment\label{const.proj}}

The method can deal with constraints, but strict constraint fulfillment is not guaranteed.
Here, we present a series of countermeasures.

Regarding equality constraints, a projection 
\begin{align*}
	w^{\rm proj}_k := \underbrace{(I-C^\T(C C^\T)^{-1}C)}_{=:K} w_k \underbrace{- C^\T ( CC^\T)^{-1} D}_{=: L} x_k 
\end{align*}
and modified control law $u_k = E w^{\rm proj}_k$ can be applied to ensure exact constraint fulfillment.
The above is the explicit solution of $\min_{w^{\rm proj}} 1/2 \norm{w^{\rm proj}-w}^2$ s.t. $h(w^{\rm proj};x)=Cw^{\rm proj}+Dx=0$.
Note that $K,L$ can be computed and stored in advance to decrease the computational burden.

If we use this projection method, the matrix $W$, that expresses the connection from $[x;u]$ to $[w;x]$, should be replaced to 
\begin{align*}
	W &= \mat{ O & I \\ EK & EL },
\end{align*}
and inequality constraint should be modified to $g(w^{\rm proj}_k)\leq 0$.
The proofs in Sections~4 do not require any correction corresponding to the above modification.

The exact fulfillment of inequality constraints is considerably more difficult.
However, we can confirm that the inequality constraints are strictly satisfied at the equilibrium point of the controller $\mathcal{C}_{\rm d}$ with fixed $x$.
\begin{thm} \label{thm_diMPC_equi}
	Under Assumptions~\ref{ass_diMPC_cont}
	, the equilibrium condition of $\mathcal{C}_{\rm d}$ for some fixed $x$ satisfies the inequality constraint $g(w)\leq 0$.
\end{thm}
\begin{pf}
	The equilibrium condition for fixed $x$, i.e., $(\Delta w, \Delta \mu, \Delta \lambda) = (0,0,0)$ yields
	\begin{align*}
		&\nabla f + \nabla g (\eta \circ \mu) + \kappa \nabla h \lambda = 0,\\
		&\eta \circ \PPPAR{g(w)}_{\mu}^+ = 0, \
			h(w;x) = \alpha \lambda .
	\end{align*}
	The condition $\eta \circ \PPPAR{g(w)}_{\mu}^+ = 0$ means satisfaction of $(g(w))_i \leq 0$ if $(\PPPAR{g(w)}_{\mu}^+)_i=0$.
	Let us consider the case if $(\eta)_i=0$.
	Since $\gamma >0$, this case is equivalent to $(\bar{\eta})_i=0 \douti (\mu)_i=0,(\PPPAR{g(w)}_{\mu}^+)_i<0$ from Definition~\ref{def_stepsize}, and this yields $(g(w))_i<0 $.
\end{pf}
This fact suggests that the speed of inequality constraint fulfillment can be adjusted by increasing the controller gain $\zeta$.
As shown in the numerical example later, the method is extremely computationally fast, so the inequality constraints can be satisfied quickly by reducing $\Delta t$ to maintain stability instead of increasing $\zeta$.

We also have other options to achieve exact inequality constraint fulfillment by employing extra optimization, e.g., projecting $w$ onto a feasible set or applying control barrier functions \cite{Ames2019}. 
However, we do not employ such methods here since our main objective is to decrease the computational load by avoiding direct optimization in the control law.

\section{Numerical Experiments}
\subsection{Control performance}

In this section, we evaluate the control performance of the proposed method using a numerical example of a DC motor, referring to the literature \cite{Yoshida2019}.
The target plant $\mathcal{P}_{\rm c}$ has one-input, two-state with matrices:
	\begin{align*}
		A_{\rm c} = \mat{-4&-0.03\\0.75&-10}, B_{\rm c} = \mat{2\\0}.
	\end{align*}
We configure a sampled-data feedback system (Figure~\ref{system}(b)) that incorporates the plant described earlier and a discrete-time PDG controller $\mathcal{C}_{\rm d}$ with various parameter settings. 
The objective is to examine the influence of the introduced gain $\zeta$ and the sampling period $\Delta t$ on control performance. 
Controller parameters are set as $(\alpha,\beta)=(0.2,0.1)$, $\gamma_k$ is determined by Algorithm~1, and equality constraint projection in Section~\ref{const.proj} is employed.
In Problem~\ref{ocp}, the control model is derived 
with the time-step $\Delta \tau = 0.1~{\rm s}$.
A target $x^r=[200/3;5]$ for the state $x$ and an upper bound $\overline{u}=160$ for the input $u$ are considered, with the objective function $f(w)= \norm{w}_P^2$, where $P = {\rm blockdiag}(I_{2N},I_{N}/10)$ and the horizon length $N=30$. 
The objective function is defined with the error system configured such that the state and input are zero at the steady state of $x=x^r,u=u^r\approx 133.4$. In other words, when the optimal solution is $w=0$, the system is in equilibrium at $x=x^r,u=u^r$.

\begin{figure}[t]
		\centering\setlength{\fboxsep}{1.5pt}
		\fbox{\scriptsize Solid: Result, Dotted: True optimal, Dashed: Reference/Bound}\vspace{5pt}
		\includegraphics[clip, scale=0.47]{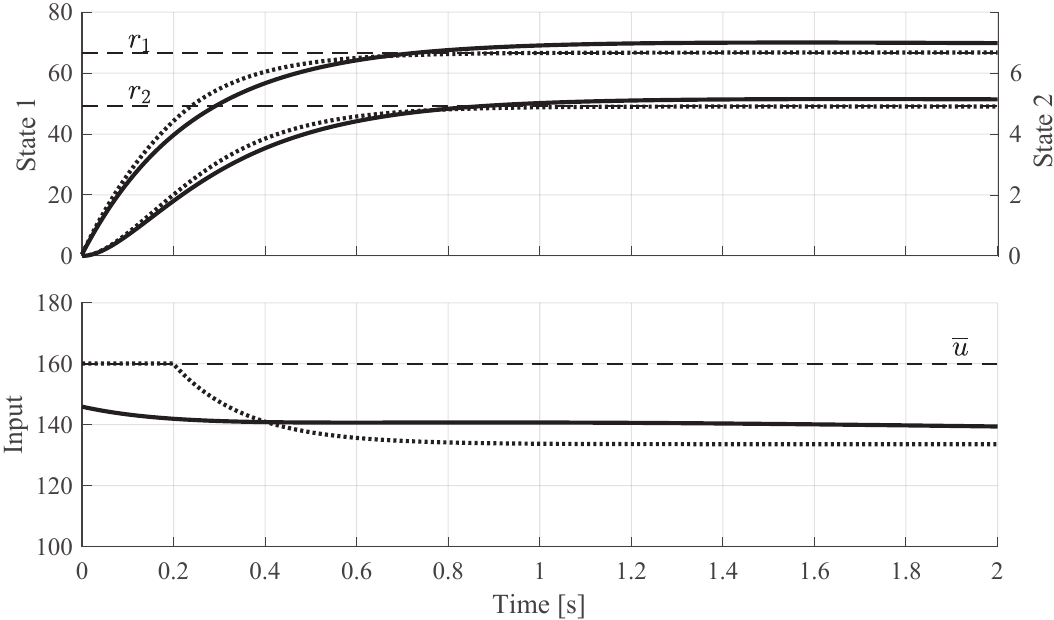}\\
		\caption{Results of Case~1 ($\zeta=1, \Delta t =$ 1~ms) \label{case1}}
	\end{figure}

\begin{figure}[t]
		\centering\setlength{\fboxsep}{1.5pt}
		\includegraphics[clip, scale=0.47]{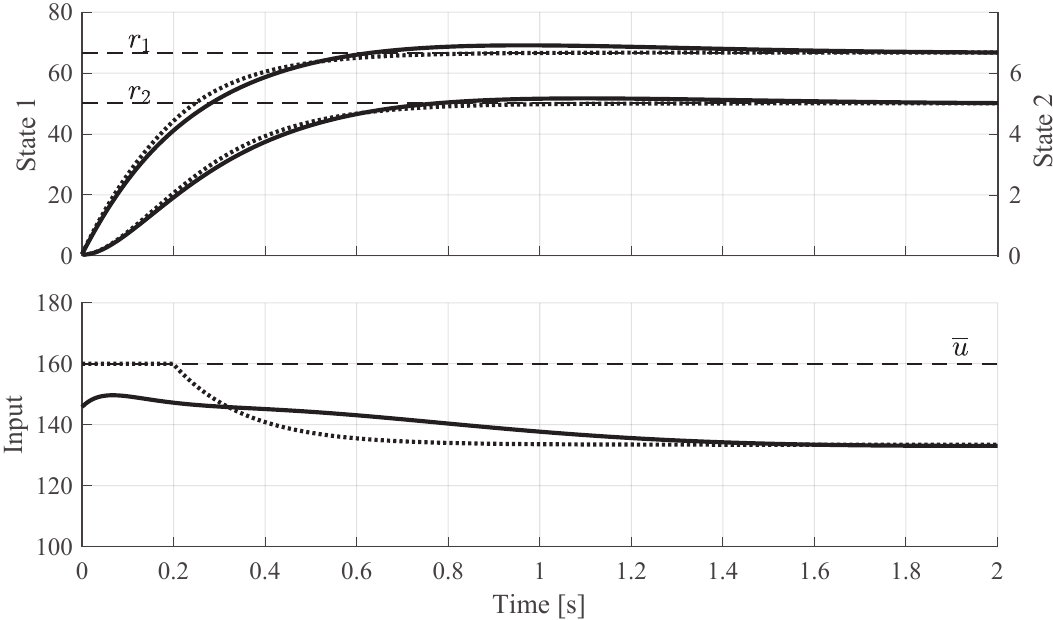}\\
		\caption{Results of Case~2 ($\zeta=10, \Delta t =$ 1~ms)\label{case2}}
	\end{figure}

\begin{figure}[t]
		\centering\setlength{\fboxsep}{1.5pt}
		\includegraphics[clip, scale=0.47]{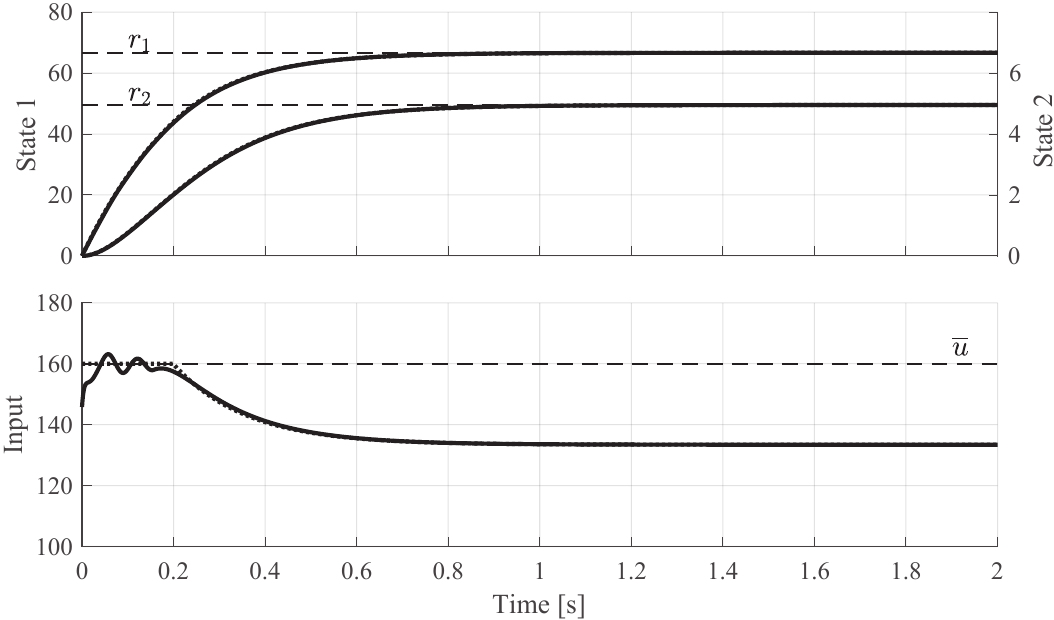}\\
		\caption{Results of Case~3 ($\zeta=100, \Delta t =$ 1~ms)\label{case3}}
	\end{figure}

\begin{figure}[t]
		\centering\setlength{\fboxsep}{1.5pt}
		\fbox{\scriptsize Solid: Result, Dotted: True optimal, Dashed: Reference/Bound}\vspace{5pt}
		\includegraphics[clip, scale=0.47]{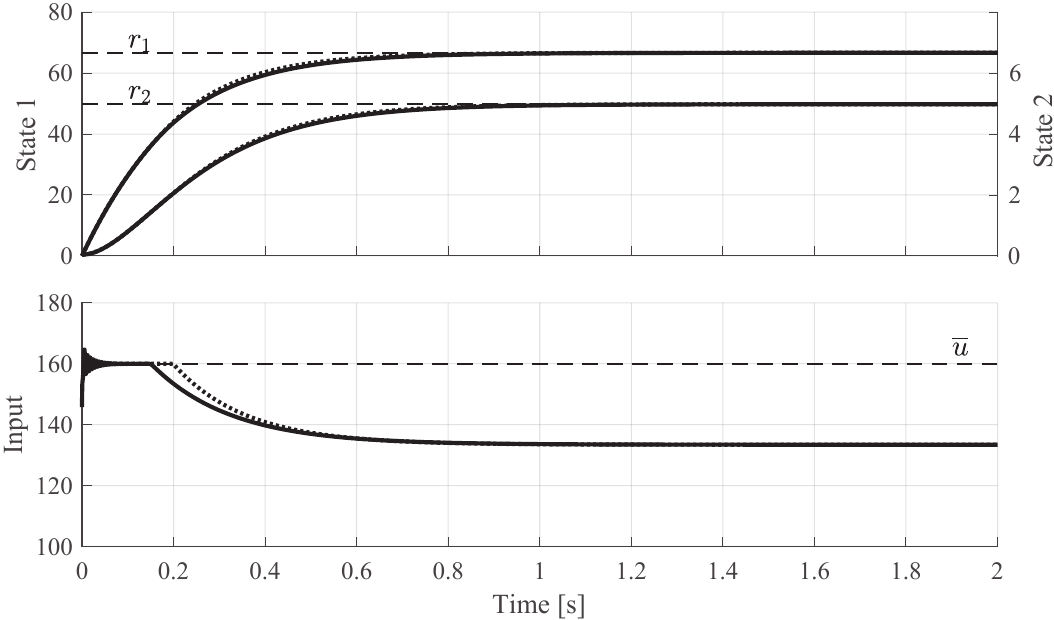}\\
		\caption{Results of Case~5 ($\zeta=1000, \Delta t =$ 0.1~ms)\label{case5}}
	\end{figure}

The case settings and the evaluated stability measures are summarized in Table~\ref{simsets}. The case numbers correspond to the controller gain $\zeta$, with larger case numbers indicating higher gains. In the table, $\delta^*_{\rm c},\delta^*_{\rm d}$ are the optimal $\delta$ values that minimize the maximum eigenvalue of $H_{\rm c}$ and $\bar{H}_{\rm d}$, obtained using an LMI solver. The resulting minimum of the maximum eigenvalue is also presented as $\lambda_{\max}^*(H_{\rm c}),\lambda_{\max}^*(\bar{H}_{\rm d})$.
The simulation results corresponding to Cases 1,2,3, and 5 are presented in Figures~\ref{case1},\ref{case2},\ref{case3} and \ref{case5}. 
In each figure, the dashed line represents the reference for the state or the upper bound for the input, and the solid line illustrates the control result for each case. 
As a baseline, the true optimal trajectory, determined by directly solving Problem~\ref{ocp} in each time step, is depicted as dotted lines.

	\begin{table}[t]
        \setlength{\tabcolsep}{4pt}\renewcommand{\arraystretch}{1.1}
		\centering
		\caption{Case settings and evaluated stability measures\label{simsets}} \vspace{5pt}
        {\scriptsize
		\begin{tabular}{c||c|c|c|c|c}
			\hline 
			Case  	 & 1 & 2 & 3 & 4 & 5\\ 
			\hline		\hline  
			$\Delta t$	& \multicolumn{4}{c|}{1~ms}  & 0.1~ms  \\ 
			\hline 
			$\zeta$ 	& 1 & 10 & 100 & \multicolumn{2}{c}{1000}\\ 
			\hline 
			$\delta^*_{\rm c}$ 	& \multicolumn{5}{c}{$0.2850$}\\ 
			$\lambda_{\max}^*(H_{\rm c})$ 	& \multicolumn{5}{c}{$-0.0870$}\\ 
			\hline 
			$\delta^*_{\rm d}$ 	& $0.3662$ & $0.3713$ & $0.4586$ & $14.7560$ & $0.4578$\\ 
			$\lambda_{\max}^*(\bar{H}_{\rm d})$ 	& $-0.0537$ & $-0.0512$ & $-0.0116$ & $3.2423$ & $-0.0116$\\ 			
			\hline 
		\end{tabular} }
	\end{table}	

In the original iMPC theory, the stability of the continuous-time feedback system $(\mathcal{P}_{\rm c}, \mathcal{C}_{\rm c})$ is guaranteed if $\lambda_{\max}^*(H_{\rm c})$ is negative. 
Consequently, the system is deemed stable in all cases due to $\lambda_{\max}^*(H_{\rm c})=-0.0870 < 0$, which is independent of $\zeta$ and $\Delta t$. 
However, as the simulations are conducted through sampled-data implementation, actual stability is compromised.
In Case~4, this resulted in an infinite loop in Algorithm~1, although even one additional iteration did not occur in other cases.
Note that the result for Case~4 diverged when Algorithm~1 was eliminated and $\gamma_k$ was fixed to 1.

By contrast, the proposed method allows for the evaluation of the impact of $\zeta,\Delta t$ on the stability of the sampled-data system through the negative definiteness of $\bar{H}_{\rm d}$. 
In fact, $\lambda_{\max}^* (\bar{H}_{\rm d})$ is positive in Case 4, indicating potential instability. 
Consequently, in Case 5, we designed $\lambda_{\max}^*(\bar{H}_{\rm d})$ to be negative by reducing $\Delta t$ with the same $\zeta$ as Case 4. 
Notably, $\delta^*_{\rm d}$ and $\lambda_{\max}^* (\bar{H}_{\rm d})$ in Case 5 are almost the same as in Case 3.
This is because $\zeta\Delta t$ exhibits the same value in both cases (see Remark~\ref{zetadeltat}).
The control results of all cases confirmed that stability is ensured and reference tracking performance is improved by increasing the gain $\zeta$.
Although it took longer for the constraint to be met in Case~3 (Figure~\ref{case3}), the controller was able to rapidly achieve the constraint by increasing $\zeta$ as seen in Case~5 (Figure~\ref{case5}). 
This behavior is attributed to the equilibrium condition of the controller satisfying the inequality constraint (refer to Theorem~\ref{thm_diMPC_equi}).

\begin{table}[t]
	\centering
	\caption{\revenv Normalized performance measures \label{p.measure}} \vspace{5pt}
	{\scriptsize \revenv
	\tabcolsep = 4pt
	\renewcommand{\arraystretch}{1.1}
	\begin{tabular}{l||cc|cc}
		\hline 
					& \multicolumn{2}{c|}{Actual} & \multicolumn{2}{c}{Horizon} \\
		Method		& obj.*1   & con.*2   & obj.*3   & con.*4 \\ 
		\hline \hline
		iMPC 		& 1.0134 & 1.1904 & 0.4111 & 187.67 \\ 
		iMPCproj 	& 1.0073 & 1	  & 1.0641 & 1	 \\ 
		C/GMRES	1	& 1.0061 & 12.515 & 1.0213 & 13.506 \\ 
		C/GMRES	2	& 1.0002 & 0.5011 & 1.0022 & 0.5453 \\ 
		MPC			& 1      & 0.0000 & 1	   & 0.0000 \\ 
		\hline 
	\end{tabular} }	\\ \vspace{5pt} \flushleft \scriptsize \noindent
	\revenv 
	The denominators for normalization are shown as "1".\\
	*1: sum of $\norm{[u-u^r;x-x^r]}^2$. *2: sum of $\norm{\max(0,u-\bar{u})}^2$. *3: sum of $f(w)$. *4: sum of $\norm{\max(0,g(w))}^2+\norm{h(w)}^2$.
\end{table}
\rev{A quantitative comparison of the control performance is shown in Table~\ref{p.measure}.
The comparison is made between the result of the proposed method in Case5 (iMPCproj), the same one without equality constraint projection (iMPC), the baseline shown as dotted lines in Figs.~\ref{case1}-\ref{case5} obtained by directly solving Problem~\ref{ocp} (MPC), and the result of C/GMRES \cite{Ohtsuka2004} (outlined in Appendix~2), which is another fast method for MPC, with one internal iteration (C/GMRES 1) and two iterations (C/GMRES 2).
As performance measures, we calculated the time summations of the objective function values and constraint violation amounts from the actual control result (Actual obj. and con.), and those from the predicted horizon at each time (Horizon obj. and con.)
The actual performance of iMPCproj was moderately better than that of iMPC, and was comparable with C/GMRES~1 for the objective function value and much better than that in constraint handling.
The performance measures for horizon show the effect of the equality constraint projection in iMPCproj, that is, the small Horizon obj. and large Horizon con. in iMPC show that the equality constraints, including the system dynamics, were not satisfied in iMPC.
We can understand that the actual performance in iMPCproj has been improved by more appropriately accounting for system dynamics.
Note that, although we tried some QP solvers and Explicit MPC to obtain the baseline (MPC) result, the measures were the same within the significant figures shown in Table~\ref{p.measure}.}

The above results support the effectiveness of the proposed method in achieving performance improvement, via introducing a controller gain and a constraint projection method, and stability guarantees while accounting for the system being implemented as a sampled-data system. 
\rev{In addition, the performance were shown to be comparable with another fast computation method that cannot ensure stability.}
Nevertheless, even with a significantly increased controller gain, the constraint violation for a very short period is inevitable. 
If one needs more stringent constraint fulfillment, some countermeasures as described at the end of Section~4.4 are necessary.

\subsection{Computational efficiency}

To evaluate the computational efficiency of the proposed method, we have considered a more practical problem regarding diesel engine airpath control \cite{Moriyasu2022}, in addition to the DC motor example shown above.
Table~\ref{cases} shows the summary of the problem sizes.
Although the proposed method can deal with a wider range of convex programs, these problems are all formulated as QPs to compare with standard linear MPC methods.

\begin{table}[t]
	\centering
	\caption{Summary of problem sizes \label{cases}} \vspace{5pt}
	{\scriptsize
	\tabcolsep = 3pt
	\renewcommand{\arraystretch}{1.1}
	\begin{tabular}{l|ccc}
		\hline 
		Dimension				& DC motor & Diesel 5 & Diesel 10 \\ 
		\hline
		State $x$					& 2 	 & 3      & 3 	 \\ 
		Input $u$					& 1 	 & 3      & 3    \\ 
		Horizon length $N$			& 30 	 & 5      & 10   \\ 	
		Primal variable	$w$			& 90 (30) & 30 (15) & 60 (30) \\ 
		Inequality constraint $g$	& 30 	 & 30     & 60 \\ 
		Equality constraint  $h$ 	& 60 (0)  & 15 (0)  & 30 (0) \\ 
		\hline 
	\end{tabular} }	\\ \vspace{5pt} \scriptsize
	The solvers other than the proposed employ a bracketed setup.
\end{table}	

We compared the proposed method with \rev{C/GMRES and} direct solution by other major QP solvers, including Explicit MPC \cite{MPC10} which creates piecewise affine control maps by solving multi-parametric QP in advance.
The comparison was made with iMPC, iMPCproj (iMPC with equality constraint projection), \rev{C/GMRES~1 (one internal iteration), C/GMRES~2 (two internal iterations),} Explicit MPC, QPKWIK \cite{qpkwik}, OSQP \cite{osqp}, and FBRS \cite{fbrs}.
Except for iMPC and iMPCproj, the state trajectory in the primal variable is eliminated by substituting the equality constraint arising from dynamics into the objective function, and the reduced problem sizes are described in parentheses in Table~\ref{cases}.
We applied a warm start to OSQP and FBRS and adjusted the convergence conditions to the same order.
All the solvers were implemented on MATLAB\raisebox{1ex}{\tiny{\textregistered}} as mex forms and executed in an Intel Core i7-1265U CPU.

\begin{table*}[t!]
	\centering
	\caption{Computation time and solver iterations \label{time}} \vspace{5pt}
	{\scriptsize
	\tabcolsep = 4pt
	\renewcommand{\arraystretch}{1.1}
	\begin{tabular}{l||ccc|cc||ccc|cc||ccc|cc}
		\hline 
			& \multicolumn{5}{c||}{DC motor} & \multicolumn{5}{c||}{Diesel 5} & \multicolumn{5}{c}{Diesel 10} \\ 
		\hline 
			& \multicolumn{3}{c|}{Time [$\mu$s]} & \multicolumn{2}{c||}{Iteration} 
			& \multicolumn{3}{c|}{Time [$\mu$s]} & \multicolumn{2}{c||}{Iteration} 
			& \multicolumn{3}{c|}{Time [$\mu$s]} & \multicolumn{2}{c}{Iteration}  \\ 
			& max & mean & iter & max & mean & max & mean & iter & max & mean & max & mean & iter & max & mean \\ 
		\hline \hline
		iMPC 		& 0.507 & $\la$ & $\la$ & 1     & 1     & 
					  0.386 & $\la$ & $\la$ & 1     & 1     &
					  0.666 & $\la$ & $\la$ & 1     & 1      \\ 
		iMPCproj 	& 1.252 & $\la$ & $\la$ & 1     & 1     &
					  0.547 & $\la$ & $\la$ & 1     & 1     &
					  1.173 & $\la$ & $\la$ & 1     & 1      \\ 
\rev{	C/GMRES~1 }	& 4.834 & $\la$ & $\la$ & 1     & 1     &
					  2.458 & $\la$ & $\la$ & 1     & 1     &
			    	  24.24 & $\la$ & $\la$ & 1     & 1     \\ 
\rev{	C/GMRES~2 }	& 6.110 & $\la$ & 1.275 & 2     & 2     &
					  3.232 & $\la$ & 0.774 & 2     & 2     &
			    	  32.14 & $\la$ & 7.906 & 2     & 2     \\ 
		Explicit MPC& 1.243 & 0.445 & 0.056 & 22    & 7.880 &
					  123.7 & 11.78 & 0.105 & 1173  & 111.7 &
					  3741  & 436.5 & 0.235 & 15917 & 1857   \\ 	
		QPKWIK 		& 19.70 & 11.67 & 6.565 & 3     & 1.778 &
					  8.489 & 2.884 & 2.122 & 4     & 1.359 &
					  44.89 & 11.61 & 6.413 & 7     & 1.810  \\ 
		OSQP(warm)	& 196.9 & 5.845 & 5.625 & 35    & 1.039 &
					  50.42 & 3.882 & 3.878 & 13    & 1.001 &
					  142.9 & 5.961 & 5.955 & 24    & 1.001  \\ 
		FBRS(warm)	& 66.08 & 9.506 & 9.440 & 7     & 1.007 &
					  42.87 & 4.763 & 4.761 & 9     & 1.000 &
					  119.8 & 13.31 & 13.81 & 9     & 1.000  \\ 
		\hline 
	\end{tabular} }	
\end{table*}

The results are summarized in Table~\ref{time}.
The table shows the average running time for one solver iteration ({\tt Time iter})\footnote{The iterations for iMPC and iMPCproj represent the while loop shown in Algorithm 1, and for the Explicit MPC mean the iterations for identifying the piecewise linear region.}, the average time for one sampling period ({\tt Time mean}), and the estimated maximum time for one sampling period ({\tt Time max}), as well as the average ({\tt Iter mean}) and maximum ({\tt Iter max}) of the number of solver iterations in one sampling period.
To eliminate the influence of unpredictable OS processes and variable CPU clock frequency, {\tt Time max} is estimated by ({\tt Time mean})$\times$({\tt Iter max})/({\tt Iter mean}).

In all problem settings, iMPC could perform one sampling period faster than only one iteration of other solvers except Explicit MPC.
While iMPC can easily take advantage of the sparse structure of the matrices, iMPCproj is more computationally demanding due to the use of dense matrices in the computation of $w^{\rm proj}$.
It showed, however, sufficiently small computation time compared with others.
Explicit MPC showed the smallest {\tt Time iter}, but as is well known \rev{in literatures such as \cite{4178103,Alessio2009}}, the number of piecewise linear regions increases explosively with the number of inequality constraints.
This resulted in a large {\tt Iter max} to identify the region and more than 60~MB of memory space to store the pre-calculated control law in the Diesel~10 problem.
Other QP solvers require solving a linear equation at each iteration, which makes {\tt Time iter} relatively large.
Although warm starting is effective for some solvers such as OSQP and FBRS to reduce {\tt Iter mean}, there is no guarantee that it will effectively reduce {\tt Iter max} in practical situations, where problem parameters, i.e., reference, bounds, etc., will be frequently changed.
\rev{In contrast, since C/GMRES fixes the number of internal iterations, there is no need to worry about variations in computation time, and the maximum run time is smaller than other QP solvers, except for Explicit MPC in DC motor case.
However, iMPC and iMPCproj are about 10 times faster than C/GMRES~1 which is the fastest setup sacrificing the performance.}

A notable advantage of iMPC against QP solvers is its iterative-free nature.
Although Algorithm~1 contains a while loop, it did not generate even one additional iteration in all sampling, and removing the loop is easy as described in Section 4.3.
One of the biggest problems in implementing a QP solver on a real system is estimating the maximum amount of time the solver will spend.
Explicit MPC, which pre-calculates a piecewise linear function, and QPKWIK, which is based on the active-set method, can estimate the upper bounds of iterations in advance, but they are often not within a realistic range because the number explodes with the size of the problem.
Other QP solvers do not guarantee to obtain a solution in the desired number of iterations.
Although stopping iterations after a fixed number is a common measure, the residuals from incomplete iterations adversely affect control performance and stability.
\rev{That is also true in C/GMRES and it additionaly has discritization errors as an inevitable error souce.}
Our proposed method allows for both reliable execution within the sampling time and guarantees stability.

\section{Conclusion}

In summary, we proposed the application of discrete-time PDG dynamics to address MPC problems and derived a sufficient condition for ensuring the stability of the resulting sampled-data feedback system. The effectiveness of our stability evaluation method was validated through simple numerical examples, demonstrating its superiority compared with conventional methods. Furthermore, our proposed method exhibited significantly faster computation times in practical examples when compared with other established approaches.

One limitation of our proposed method is that strict inequality constraint fulfillment is not inherently guaranteed. 
A simple countermeasure would be to use a projection onto the feasible set or a control barrier function \cite{Ames2019}, but this would introduce an additional optimization problem in general and would lose the advantages of our method.
In the context of approximate MPC, a method has been proposed to switch to a pre-designed control law that satisfies the constraints when the approximation of the optimal solution does not satisfy the constraints \cite{hose2023approximate}, which might also be used effectively in our method.
Another issue is that the stability assurance of this method heavily depends on the dissipative nature of the plant and sometimes includes large conservatism, and thus it may be challenging for certain types of plants to guarantee stability or to achieve high control performance. 
Future work will focus on resolving these issues and extending the theory and control system to enhance their universality and applicability across a wider range of scenarios.

\bibliographystyle{plain}        
\bibliography{MyCollection}           

\appendix
\section*{Appendix~1 \ Proof of Lemma~\ref{diMPC_cont}\label{app.lemma}}
\begin{pf}
	By utilizing $(\nabla f(w))^\T w \geq \sigma w^\T w$, which stems from Assumption~\ref{ass_iMPC2} (a), and Assumption~\ref{ass_iMPC2} (c), we can show
	\footnotesize \vspace{-10pt}
	\begin{align*}
		\Delta S_k^{\rm I} &:= w_k^\T \Delta w_k + \mu_k^\T\Delta \mu_k + \lambda_k^\T\Delta \lambda_k \\
		&= \zeta \Delta t \left( -w_k^\T (\nabla f + \tau\kappa\nabla h(\lambda_k+\beta h) ) \right. \\
			& \ \ \ \ \ \ \ \ \ \ \,
				 + \underbrace{(\bar{\eta}_k \!\circ\! \mu_k)^{\!\T} \PAR{[g(w_k)]_{\mu_k}^+ \!-\! (\nabla g)^{\!\T} \! w_k \!} }_{=:b^{\rm I}(\mu_k,z_k)} \gamma_k \\
			& \ \ \ \ \ \ \ \ \ \
				\left. + \, \tau \lambda_k^\T (-\alpha \lambda_k + h) \right)\\
		&\leq \zeta \Delta t \left( -w_k^\T (\rho w_k \!+\! \tau\kappa C^\T (\lambda_k \!+\! \beta Cw_k \!+\! \beta Dx_k) ) \right. \\
			& \ \ \ \ \ \ \ \ \ \ \left.
				+ \, \tau \lambda_k^\T (-\alpha \lambda_k \!+\! Cw_k \!+\! Dx_k) \!+\! b^{\rm I}(\mu_k,z_k) \gamma_k \right)\\
		&= \zeta \Delta t \PAR{ z_k^\T \bar{H}^{\mathcal{C}}_{\rm c} z_k + b^{\rm I}(\mu_k,z_k) \gamma_k }.
	\end{align*}\normalsize
	In addition, Assumption~\ref{ass_iMPC2} (c) and Lemma~\ref{lem_smooth} yields \footnotesize \vspace{-10pt}
	\begin{align*}
		\Delta S_k^{\rm II} &:= \frac{1}{2} \PAR{ \Delta w_k^\T \Delta w_k \!+\! \Delta \mu_k^\T \Delta \mu_k 
			\!+\! \Delta \lambda_k^\T \Delta \lambda_k} \\
		& = (\zeta \Delta t)^2 \left( \frac{1}{2}\norm{\nabla f + \tau \kappa \nabla h \PAR{\lambda_k + \beta h(w_k;x_k) }}^2 \right.\\
			&\ \ \ \ \ \
				+ \underbrace{(\bar{\eta}_k\!\circ\! \mu_k)^\T \PAR{ \PAR{\nabla g}^\T \PAR{\nabla f \!+\! \tau \kappa \nabla h \PAR{\lambda_k \!+\! \beta h }} } }_{=: b^{\rm II}(\mu_k,z_k)} \gamma_k \\
			&\ \ \ \ \ \
				+ \underbrace{\frac{1}{2} \PAR{ \norm{\bar{\eta}_k \circ [g(w_k)]_{\mu_k}^+ }^2 + \norm{\nabla g (\mu_k \circ \bar{\eta}_k)}^2 }}_{=: a^{\rm II}(\mu_k,z_k)} \gamma_k^2 \\
			& \ \ \ \ \ \ \left. 
				+ \frac{\tau^2}{2} \norm{Cw_k + D x_k -\alpha \lambda_k }^2\right) \\
		& \leq (\zeta \Delta t)^2 \left( \frac{1}{2} z_k^\T \bar{P} z_k + \frac{\tau^2}{2} \norm{ \PPPAR{C \ D \ -\alpha I} z_k }^2 \right.\\
			& \ \ \ \ \ \ \ \ \ \ \ \ \ \ \left. 
				+\, b^{\rm II}(\mu_k,z_k) \gamma_k + a^{\rm II}(\mu_k,z_k) \gamma_k^2\right) \\
		& = (\zeta \Delta t)^2 \left( z_k^\T{\bar{P}^{\mathcal{C}}_{\rm d}}z_k \!+\! a^{\rm II}(\mu_k,z_k) \gamma_k^2 \!+\! b^{\rm II}(\mu_k,z_k) \gamma_k \right).
	\end{align*}\normalsize
	Since $\Delta S^{\mathcal{C}}_k=\Delta S_k^{\rm I}+\Delta S_k^{\rm II}$, adding the above two inequalities together yields the claimed inequality with $a(\mu,z) =\zeta \Delta t a^{\rm II}(\mu,z), \, b(\mu,z) = b^{\rm I}(\mu,z)+\zeta \Delta t b^{\rm II}(\mu,z)$. 
	The claimed properties $a(0,0)=0, b(0,z) = 0 \ \forall z$ hold trivially and $b(\mu,0)<0 \ \forall \mu \neq0$ can be confirmed from $\nabla f(0) =0, g(0)<0$ in Assumption~\ref{ass_iMPC2} (a) and (b), respectively.
\end{pf}

\section*{\rev{Appendix~2 \ Control law of C/GMRES}\label{cgmres}}
\revenv
For simplicity, we consider the case without equality constraints.
The first-order optimality condition for a problem $\minimize_w \ f(w;t) \ {\rm s.t.} \ g(w,t) \leq 0$ is:
\begin{align*}
	& \nabla f(w;t) + \nabla g(w;t) \nu = 0, \\ 
	& \nu \geq 0, \ g(w;t) \leq 0, \ \nu^\T g(w;t) = 0,
\end{align*}
where $\nu$ is a Lagrange multiplier. 
If $h$ in Problem~\ref{ocp} only includes the system dynamics, we can reduce the problem into the above. 
With some complementarity function $\phi$, the above condition is equivalent to
\begin{align*}
	F(\omega;t) := \mat{\nabla f(w;t) + \nabla g(w;t) \nu \\ \phi(-g(w;t), \nu)} = 0, 
\end{align*}
where $\omega:=[w;\nu]$.
We employed Fisher-Burmeister function $\phi(a,b) = a + b - \sqrt{a^2 + b^2}$ in this paper.

C/GMRES defines a dynamics $\dot{F} = -\xi F$ with a positive constant $\xi$, as a control law, and this yields
\begin{align*}
	\pdiff{F}{\omega} \dot{\omega} = -\PAR{\xi F + \pdiff{F}{t}},
\end{align*}
where $\del F/\del t$ typically contains the system dynamics' effect.
The above is just a linear equation and the time-derivative of control input can be determined by solving it.
iMPC can be regarded as the method which ignores the relationship among variables and approximates the above control law into a decoupled one.
From this idea, we employed $\xi = \zeta$ in this paper.

C/GMRES applies generalized minimal residual method (GMRES) to solve it and typically limits its iteration to be a small number to achieve fast computation.
Note that, since there are some discretization errors and residuals, the stability and constraint fulfillment are not guaranteed in practice.

\end{document}